\documentclass[aps,prd,groupedaddress,preprint,eqsecnum,nofootinbib]{revtex4}
\usepackage{graphicx,epsf,amssymb,amsbsy,amsfonts,amssymb,amsmath,physics,courier,hyperref,ytableau,xcolor}
\def\be{\begin{equation}}
	\def\ee{\end{equation}}
\def\bea{\begin{eqnarray}}
	\def\eea{\end{eqnarray}}

\def\bg{\bar{g}}
\def\beq{\begin{eqnarray}}\def\eeq{\end{eqnarray}}
\def\ba#1\ea{\begin{align}#1\end{align}}
\def\bg#1\eg{\begin{gather}#1\end{gather}}
\def\bm#1\em{\begin{multline}#1\end{multline}}
\def\bmd#1\emd{\begin{multlined}#1\end{multlined}}

\def\D{\Delta}

\def\({\left(}
\def\){\right)}
\def\[{\left[}
\def\]{\right]}

\def\D{\Delta}

\begin{document}
	\hfuzz 9pt
	\title{Note on higher spins and holographic symmetry algebra}
	
	\author{Shamik Banerjee$^{1,2}$}\email{banerjeeshamik.phy@gmail.com}
	\author{Suman Guchait$^{1,2}$}\email{sumanguchait1998@gmail.com}
	\author{Raju Mandal$^{1,2,3}$}\email{rajuphys002@gmail.com} 
	\author{Sudhakar Panda$^{1}$}\email{panda@niser.ac.in}
	\affiliation{$^1$National Institute of Science Education and Research (NISER), Bhubaneswar 752050, Odisha, India}
	\affiliation{$^2$Homi Bhabha National Institute, Training School Complex, Anushakti Nagar, Mumbai 400094, India}
	\affiliation{$^3$Indian Institute of Technology Ropar, Punjab 140001, India}

	\begin{abstract} 
	In this paper we discuss a higher spin extension of the holographic symmetry algebra for graviton and gluon. Our primary observation is that in the presence of higher spin particles the soft symmetry algebra has a subalgebra isomorphic to $w_{\infty}$ which is generated by the \textit{conformally soft higher spin particles}. This $w_{\infty}$ subalgebra does not commute with the $w_{1+\infty}$ subalegbra generated by the conformally soft gravitons. The same thing holds for the colored higher spin particles. One gets a subalgebra isomorphic to the $S$-algebra which is generated by the conformally soft colored higher spin particles. We further verify the soft algebra for colored higher spin particles using the (tree-level) $4$-point MHV amplitude of the higher spin Yang-Mills theory constructed in arXiv:2210.07130. At the end we also discuss the higher spin extension of the deformed holographic symmetry algebra for non-zero cosmological constant as constructed in arXiv:2312.00876.
		\end{abstract}
	
	\maketitle
	\tableofcontents
	\flushbottom
	
	\section{Introduction}
    \vspace{-4mm}

	Celestial holography \cite{Pasterski:2021rjz,Donnay:2023mrd,Pasterski:2016qvg,Banerjee:2018gce} is known to be very useful for studying symmetry properties of scattering amplitudes \cite{Strominger:2017zoo,Strominger:2013lka,Barnich:2009se,Pate:2019lpp,Banerjee:2020zlg,Guevara:2021abz,Strominger:2021mtt,Himwich:2023njb,Himwich:2021dau,Adamo:2021lrv,deGioia:2023cbd,Himwich:2025ekg,Sheta:2025oep} in asymptotically flat space-time. One of the central results in Celestial Holography is that the conformally soft gravitons generate a $w_{1+\infty}$ algebra \cite{Strominger:2021mtt} which is a symmetry of a certain class of graviton scattering amplitudes. For example, tree level MHV graviton scattering amplitudes and amplitudes of the selfdual gravity both have $w_{1+\infty}$ symmetry \cite{Ball:2021tmb}. The story of gluon goes along the same line. Conformally soft gluons generate an infinite dimensional $S$-algebra \cite{Strominger:2021mtt} and both tree level MHV amplitudes and amplitudes of selfdual Yang-Mills are invariant under this symmetry. So it is natural to ask what happens if we have massless higher spin particles \cite{Sorokin:2004ie} in the bulk. In asymptotically flat space-time there is no known (local and unitary) interacting theory of higher spin particles. However, in recent times examples of higher spin extension of selfdual Yang-Mills and selfdual gravity in asymptotically flat space-time have been constructed in \cite{Ponomarev:2016lrm,Ponomarev:2017nrr,Krasnov:2021nsq}. Higher spin extension of the Yang-Mills theory has also been constructed in \cite{Adamo:2022lah}. So it is natural to wonder how they fit into the Celestial holography picture. First steps in this direction were taken in the papers \cite{Himwich:2021dau,Monteiro:2022xwq}. Present paper is a step in similar direction. We present a higher spin extension of the holographic symmetry algebra obtained by analytically continuing the (helicity) spin of the particle. Our primary observation is that in the presence of higher spin particles the soft symmetry algebra has a subalgebra isomorphic to $w_{\infty}$ which is generated by the \textit{conformally soft higher spin particles}. This $w_{\infty}$ subalgebra does not commute with the $w_{1+\infty}$ subalegbra generated by the conformally soft gravitons. The same thing holds for the colored higher spin particles. One gets a subalgebra isomorphic to the $S$-algebra which is generated by the conformally soft colored higher spin particles. We further verify the soft algebra for colored higher spin particles using the (tree-level) $4$-point MHV amplitude of the higher spin Yang-Mills theory constructed in \cite{Adamo:2022lah}. It is very likely that the higher spin algebras that we discuss is actually a symmetry of the selfdual theories constructed in \cite{Ponomarev:2016lrm,Ponomarev:2017nrr,Krasnov:2021nsq}. It will be very interesting to prove this. At the end we discuss the higher spin extension of the deformed holographic symmetry algebra for non-zero cosmological constant as constructed in \cite{Taylor:2023ajd,Sheta:2025oep,Strominger:2026yrh} \footnote{See also \cite{Bittleston:2024rqe} for recent work on AdS$_4$ deformation of celestial symmetries.}. This extension is also obtained by analytic continuation in the spin of the particle. 
	
	Last but not the least we would like to mention that there is a different approach to higher spin holography in asymptotically flat space-time which has been explored in \cite{Ponomarev:2022ryp,Ponomarev:2022qkx}. It will be very interesting to compare and contrast this approach with the Celestial holography approach which we take in the present paper. We hope to return to this in future.

	\section{Higher spin OPE and symmetry algebra}\label{sec:hsym}
In Celestial holography the OPE between two positive helicity gravitons is known to be given by \cite{Guevara:2021abz}
	\be
	\label{eq:GG ope}
	G_{h_1,\bar h_1}(z_1,\bar z_1) G_{h_2,\bar h_2}(z_2,\bar z_2) \sim - \frac{\kappa}{2} \frac{1}{z_{12}} \sum_{n=0}^{\infty} B(2\bar h_1 +1 + n, 2\bar h_2 +1) \frac{\bar z_{12}^{n+1}}{n!} \bar\partial_2^{n} G_{h_1+h_2-1, \bar h_1 +\bar h_2 +1}(z_2,\bar z_2)
	\ee
	
	where $(h,\bar h)$ are the $SL(2,\mathbb{R})_L \times SL(2,\mathbb{R})_R$ weights of the graviton primary $G_{h,\bar h}(z,\bar z)$. The OPE coefficient of the leading term of $\mathcal{O}(\frac{\bar z}{z})$ is fixed by the known collinear factorization formula for two positive helicity gravitons. The OPE coefficients of the subleading terms of $\mathcal{O}(\frac{\bar z^n}{z})$, $n>1$, are completely fixed by the $SL(2,\mathbb{R})_R$ invariance of the OPE. This is reflected in the fact that the all the OPE coefficients are naturally functions of the right scaling dimension $\bar h$. 
	
	It is clear that the OPE \eqref{eq:GG ope} remains Poincare invariant for arbitrary choice of $(h,\bar h)$. In other words, if we analytically continue in $(h,\bar h)$ then we get a valid OPE for every choice of $(h,\bar h)$. This is true in spite of the fact that originally \eqref{eq:GG ope} was derived for two positive helicity gravitons for which $h-\bar h=2$. Now in celestial holography the dimension $\Delta = h+\bar h$ of a conformal primary is usually analytically continued off the principal series representation for which $\Delta= 1+ i\lambda, \lambda \in \mathbb{R}$. However, the helicity of the particle, which is given by $\sigma=h-\bar h$, remains fixed. But if we vary $(h,\bar h)$ freely, which the OPE \eqref{eq:GG ope} allows us to do, we also change the helicity of the particle. So we can also get a Celestial OPE for higher spin particles in the bulk starting from \eqref{eq:GG ope} and it has the following structure:
	
	\be\label{HSO}
	G^{\sigma_1}_{\D_1}(z_1,\bar z_1) G_{\D_2}^{\sigma_2}(z_2,\bar z_2) \sim - \frac{\kappa}{2}\frac{1}{z_{12}} \sum_{n=0}^{\infty} B(\D_1 -\sigma_1 +1 + n, \D_2 -\sigma_2 +1) \frac{\bar z_{12}^{n+1}}{n!} \bar\partial_2^{n} G^{\sigma_1 +\sigma_2 -2}_{\D_1 +\D_2}(z_2,\bar z_2)
	\ee
	
	Here we have used a slightly different notation for the celestial primary and $G^{\sigma}_{\Delta}$ denotes a primary of dimension $\D$ and helicity $\sigma$. We take the helicity $\sigma$ to be a positive integer and restrict $\sigma\ge 2$. The spin selection rule $G^{\sigma_1}\times G^{\sigma_2}\sim G^{\sigma_1+\sigma_2-2}$ which follows from the OPE \eqref{HSO} ensures that the graviton primaries form a closed algebra. This is important because we still want the Poincare algebra $ISO(3,1)$ to be a subalgebra of the higher spin symmetry algebra. Secondly, the spin selection rule implies that if we introduce a spin-$3$ particle in the bulk then we have to introduce all spins ranging from $4$ to $\infty$ otherwise the OPE will not close. This follows from the relation $G^{3}\times G^{\sigma}\sim G^{\sigma+1}$. So the spin-$3$ primary acts as a spin raising operator. 
	
	We would like to point out that the reason we do not have a spin-$1$ particle in the bulk is because a spin-$1$ particle acts a spin lowering operator which follows from the selection rule $G^{1}\times G^{\sigma} \sim G^{\sigma-1}$. So if there is a spin-$1$ particle in the bulk then we must have particles of all spin ranging from $-\infty$ to $\infty$ otherwise the OPE will not close. But this immediately leads to a contradiction. The OPE \eqref{HSO} is not valid for example, for two negative helicity gravitons. Similarly we cannot have a spin-$0$ particle in the bulk because that will require the presence of a spin-$1$ particle if we want the OPE to close. So the OPE makes sense only for $\sigma\ge 2$ and the interacting higher spin theory in the bulk from which the OPE arises must be chiral. 
	
	Now in \cite{Himwich:2021dau}, following \cite{Pate:2019lpp}, a more general higher spin OPE was written down whose spin selection rule is $G^{\sigma_1}\times G^{\sigma_2}\sim G^{\sigma_1+\sigma_2 -p -1}$ where $p$ is an integer. The OPE \eqref{HSO} corresponds to the special case $p=1$. As we will show in this paper the higher spin symmetry algebra for $p=1$ has a richer structure and one gets a $w_{1+\infty}$ algebra from bulk gravitons and another $w_{\infty}$ algebra from all the bulk higher spin particles. 
		
\subsection{Symmetry algebra}	
		
	To find the (soft) symmetry algebra we define the higher spin conformally soft \cite{Donnay:2018neh,Guevara:2021abz} operators as
	
	\be
	\label{eq:soft graviton definition}
	H^{k,\sigma}(z,\bar z) = \lim_{\D\rightarrow k} (\D-k) G_{\D}^{\sigma} (z,\bar z),
	 ~~~ k=\sigma, \sigma-1, \sigma-2,\cdots 
	\ee
	
	with weights
    \begin{equation}
    	\label{eq:weights}
    	\left( h, \bar h \right)= \left( \frac{k+\sigma}{2}, \frac{k-\sigma}{2}\right).
    \end{equation}
    
    It follows from \eqref{eq:GG ope} that the OPE between two conformally soft operators is given by
		\begin{equation}
		\begin{gathered}
			H^{k,\sigma_1}(z_1,\bar z_1)H^{l,\sigma_2}(z_2,\bar z_2)\\
			\sim -\frac{\kappa}{2}\frac{1}{z_{12}}\sum_{p=0}^{\sigma_1-k-1}\frac{(-k-l+\sigma_1+\sigma_2-2-p)!}{(-k+\sigma_1-p-1)!(-l+\sigma_2-1)!}\frac{\bar z_{12}^{p+1}}{p!}\bar \partial_2^pH^{k+l,\sigma_1+\sigma_2-2}(z_2,\bar z_2)
			\label{softsoft}
		\end{gathered}
	\end{equation}

	Now we do the truncated mode expansion \cite{Banerjee:2020zlg,Guevara:2021abz} of the soft operator $H^{k,\sigma}(z,\bar z)$ in $\bar z$ as
	
	\be
	\label{eq:mode expansion}
	H^{k,\sigma}(z,\bar z) = \sum_{n=\frac{k-\sigma}{2}}^{\frac{\sigma -k}{2}} \frac{H^{k,\sigma}_n(z)}{\bar z^{n + \frac{k-\sigma}{2}}}
	\ee
	
	and define the global modes of the currents $H^{k,\sigma}_n(z)$ as 
	
	\be
	H^{k,\sigma}_n = \oint_{z=0} dz H^{k,\sigma}_n (z)
	\ee
	
 The algebra of these generators which follow from \eqref{softsoft} is given by
	
	\be
	\begin{gathered}
		\[ H^{k,\sigma_1}_{m}, H^{l,\sigma_2}_{n}\] \\ = - \frac{\kappa}{2} \[ n(\sigma_1 -k) - m(\sigma_2-l)\] \frac{\( \frac{\sigma_1-k}{2} -m + \frac{\sigma_2-l}{2} - n -1\)!}{\( \frac{\sigma_1-k}{2} -m \)! \( \frac{\sigma_2-l}{2} - n\)!}\frac{\( \frac{\sigma_1-k}{2} +m + \frac{\sigma_2-l}{2} +n -1\)!}{\( \frac{\sigma_1-k}{2} +m \)! \( \frac{\sigma_2-l}{2} +n\)!} H^{k+l, \sigma_1+\sigma_2-2}_{ m+n}
	\end{gathered}
	\ee
	
	Now we define the set of light transformed operators \cite{Strominger:2021mtt}
	
	\be
	T^{p,\sigma}_m = \frac{1}{\kappa} (p-1 - m)! (p-1 + m)! H^{-2p+2+\sigma, \sigma}_m
	\ee
	
	whose algebra is given by 
	
	\be\label{hsa}
	\[ T^{p,\sigma_1}_m , T^{q,\sigma_2}_n \] = \[m(q-1) - n(p-1)\] T^{p+q-2, \sigma_1 +\sigma_2 -2}_{m+n}
	\ee
	where $p = 1, \frac{3}{2}, 2,....$ and $1-p \le m \le p-1$. The algebra \eqref{hsa} is the simplest possible higher spin generalization of the (wedge subalgebra of the) $w_{1+\infty}$ algebra which arises when there is only graviton in the bulk \cite{Strominger:2021mtt,Himwich:2023njb,Adamo:2021lrv}. \\
	
	\textit{It is very likely that the algebra \eqref{hsa} is an exact symmetry of the higher spin extension of the selfdual gravity theory constructed in \cite{Ponomarev:2016lrm, Ponomarev:2017nrr, Krasnov:2021nsq}. It will be really interesting to prove this.} \\

	We now describe some interesting subalgebras of the higher spin algebra \eqref{hsa}. 
	
	\section{Higher spin extension of the Poincare algebra}
	
	In the presence of higher spin particles the Poincare symmetry gets enhanced to an infinite dimensional higher spin symmetry which we now discuss. 
	
	\subsection{Higher spin translation}
	
	In a theory of gravity the conformally soft operator $H^{1,2}(z,\bar z)$ is the generator of supertranslation. Similarly the conformally soft higher spin operator $H^{\sigma-1,\sigma}(z,\bar z)$ can be thought of as the generator of spin-$\sigma$ supertranslation. This is also supported by the fact that the global modes $H^{\sigma-1,\sigma}_n$ commute, i.e, 
	\be
	\[ H^{\sigma-1,\sigma}_m, H^{\sigma'-1,\sigma'}_{m'}\] = 0, \ m, m' = \pm\frac{1}{2}
        \ee
        The action of the the global current $H^{\sigma-1,\sigma}(\bar z)$ defined as, 
        \be
        H^{k,\sigma}(\bar z)= \oint_{z=0} dz H^{k,\sigma}(z,\bar z)
        \ee
        acts on a hard higher spin operator as 
        \be\label{action}
        \[ H^{\sigma-1,\sigma}(\bar z_1), G^{\sigma'}_{\Delta}(z_2,\bar z_2)\] = \frac{\kappa}{2} \bar z_{21} G^{\sigma+\sigma'-2}_{\D + \sigma -1}(z_2,\bar z_2)
        \ee
        So the higher spin translations raise the conformal dimension $\D$ of an operator and also changes the spin. This is unlike the usual space-time translations coming from the spin-$2$ graviton and is related to the fact that the higher spin algebra closes only if we have particles of all spins starting from spin-$3$. The action on a hard particle given in \eqref{action} closely resembles the action proposed in \cite{Campoleoni:2017mbt} except that now the translation also shifts the spin of the particle. 
        
        \subsection{Higher spin analog of Lorentz transformations}
        
        Similar to translation the Lorentz group also gets an infinite dimensional extension in the presence of higher spin particle. The global modes $H^{\sigma-2,\sigma}_n$, $n=0,\pm 1$, of the soft operators $H^{\sigma-2,\sigma}(z,\bar z)$ for all $\sigma\ge 2$ form a closed algebra which may be called the higher spin extension of the Lorentz algebra. The algebra is given by 
        \be
        \begin{gathered}
        \[ H^{\sigma-2,\sigma}_m, H^{\sigma'-2,\sigma'}_{m'} \] 
        = \kappa (m-m') \frac{(1-m-m')!}{(1-m)!(1-m')!}\frac{(1+m+m')!}{(1+m)!(1+m')!} H^{\sigma+\sigma'-4, \sigma+\sigma'-2}_{m+n}
        \end{gathered}
        \ee
        
        The higher spin Lorentz generators and the higher spin translation generators form a closed algebra as one can easily verify that 
        \be
        \[ H^{\sigma-2,\sigma}_m, H^{\sigma'-1,\sigma'}_n\] \sim H^{\sigma+\sigma'-3, \sigma+\sigma' -2}_{m+n}
        \ee
	So the structure of the higher spin extension of the Poincare algebra is the same as the spin-$2$ Poincare algebra in the sense that it is again a semi-direct product of the higher-spin Lorentz algebra and the higher spin translation algebra. Just like the usual translations the higher spin translations form an Abelian invariant subalgebra of the higher spin Poincare algebra. 
	
	\section{Two copies of w-infinity}
	
	\subsubsection{$w_{1+\infty}$ from conformally soft graviton}
	
	The most obvious infinite dimensional subalgebra is the $w_{1+\infty}$ generated by the bulk gravitons \cite{Strominger:2021mtt}. This is the algebra of the set of generators $\{ T^{p,2}_m\}$, 
	
	\be
	\[ T^{p,2}_m , T^{q,2}_n \] = \[m(q-1) - n(p-1)\] T^{p+q-2, 2}_{m+n}
	\ee
	
	
	The set of generators $\{T^{p,\sigma}_m\}$ with a \textit{fixed} value of $\sigma \ge 3$ transform in the adjoint representation of this $w_{1+\infty}$ algebra as can be seen from the commutation relation
	
	\be
	\[ T^{p,2}_m , T^{q,\sigma}_n \] = \[m(q-1) - n(p-1)\] T^{p+q-2, \sigma}_{m+n}
	\ee
	
	\subsubsection{$w_{\infty}$ from conformally soft higher spin particles}
	
	Let us consider the following subset of generators defined as 
	
	\be
	\tilde w^{p}_m = T^{p,2p-2}_m, \ p= 2, 5/2,\cdots
	\ee
	
	The restriction on $p$ follows from the fact that we do not have spin-$1$ particle in the bulk. It is easy to see that this set of generators form another $w_{\infty}$ algebra 
	
	\be
	\[ \tilde w^{p}_m , \tilde w^{q}_n \] = \[m(q-1) - n(p-1)\] \tilde w^{p+q-2}_{m+n}, \  p=2,\frac{5}{2},\cdots
	\ee
	
	It will be very interesting to reproduce this $w_{\infty}$ algebra from an asymptotic symmetry analysis in four dimensional asymptotically flat space-time. 
	
	The appearance of two non-commuting copies of $w$-algebras is reminiscent of the higher-spin square structure found by the authors \cite{Gaberdiel:2017ede} in the case of tensionless string theory on $AdS_3\times S^3\times T^4$. It will be very interesting to clarify this connection further.

	\section{Higher spin generalization of the $S$-algebra and Higher spin Yang-Mills (HSYM)}
	
	In this section we write down the higher spin generalization of the $S$-algebra for gluons. We assume that all the coloured higher spin particles have helicities $\ge 1$. The celestial OPE between two positive helicity outgoing gluons is given by \cite{Guevara:2021abz}
	\begin{equation}
		\label{eq:hs1}
		\begin{gathered}
			\mathcal{O}^{a}_{h_1,\bar{h}_1}(z_1,\bar{z}_1)\mathcal{O}^{b}_{h_2,\bar{h}_2}(z_2,\bar{z}_2)\\
			\sim -\frac{if^{abc}}{z_{12}}
			\sum_{m=0}^{\infty}
			B(2\bar{h}_{1}+n,2\bar{h}_{2})
			\frac{\bar{z}^{m}_{12}}{m!}\bar{\partial}^{m}\mathcal{O}^{c}_{h_1+h_2-1,\bar{h}_{1}+\bar{h}_{2}}(z_{2},\bar{z}_{2})
		\end{gathered}
	\end{equation}
	where $f^{abc}$ are structure constants of the gauge group and $B(x,y)=\frac{\Gamma(x)\Gamma(y)}{\Gamma(x+y)}$ is the Euler beta function. To get the simplest higher spin extension of this OPE we can analytically continue in $(h,\bar h)$. The resulting OPE between two outgoing colored particles of arbitrary spins is given by
	\begin{equation}
		\label{eq:hs2}
		\begin{gathered}
			\mathcal{O}^{\sigma_1,a}_{\Delta_1}(z_1,\bar{z}_1)\mathcal{O}^{\sigma_2,b}_{\Delta_2}(z_2,\bar{z}_2)\\
			\sim
			-\frac{if^{abc}}{z_{12}}
			\sum_{m=0}^{\infty}
			B(\Delta_1+n-\sigma_1,\Delta_2-\sigma_2) 
			\frac{\bar{z}^{m}_{12}}{m!}
			\bar{\partial}^{m}\mathcal{O}^{\sigma_1+\sigma_2-1,c}_{\Delta_1+\Delta_2-1}(z_{2},\bar{z}_{2})
		\end{gathered}
	\end{equation}
	where $\sigma$
 is the helicity of the particle.	
 
 	Now conformally soft operators \cite{Donnay:2018neh,Guevara:2021abz} can be defined as
	\begin{equation}
		\label{eq:soft1}
		R^{k,\sigma,a}(z,\bar{z})=\lim_{\Delta \rightarrow k} (\Delta-k)\mathcal{O}^{\sigma,a}_{\Delta}(z,\bar{z}),
		~~~k=\sigma, \sigma-1,\sigma-2,\cdots 
	\end{equation} 
	with weights
	\begin{equation}
		\label{eq: soft gluon weights}
		\left( h, \bar h \right)= \left( \frac{k+\sigma}{2}, \frac{k-\sigma}{2}\right).
	\end{equation}
	
	The OPE between two soft operators is given by
	\begin{equation}
		\label{eq:hs3}
		\begin{gathered}
			 R^{k,\sigma_1,a}(z_1,\bar{z}_{1})R^{l,\sigma_2,b}(z_{2},\bar{z}_{2})\\
			 \sim
			-\frac{if^{abc}}{z_{12}}
			\sum_{m=0}^{\sigma_1-k} \frac{1}{(\sigma_1-k-m)!}
			\frac{(\sigma_1+\sigma_2-k-m-l)!}{(\sigma_2-l)!}
			\frac{\bar{z}^{m}_{12}}{m!}
			\bar{\partial}^{m}R^{k+l-1,\sigma_1+\sigma_2-1,c}(z_2,\bar{z}_{2})
		\end{gathered}
	\end{equation}
	The structure of the Beta function in the OPE \eqref{eq:hs2} allows us to define the following truncated mode expansion of $R^{k,\sigma,a}(z,\bar{z})$ in the $\bar{z}$ variable
	\begin{equation}
		\label{eq:hs gluon modes}
		R^{k,\sigma,a}(z,\bar{z})
		=\sum_{n=\frac{k-\sigma}{2}}^{\frac{\sigma-k}{2}}
		\frac{R^{k,\sigma,a}_{n} (z)}{\bar{z}^{n + \frac{k-\sigma}{2}}}.
	\end{equation}
	The global modes of the currents $R^{k,\sigma,a}_{n} (z)$, defined as,
	\be
	R^{k,\sigma,a}_{n} = \oint_{z=0} dz R^{k,\sigma,a}_{n} (z)
	\ee
	satisfy the following commutation relation,
	\begin{equation}
		\label{eq:hs4}
		\begin{gathered}
			[R^{k,\sigma_1,a}_{n},R^{l,\sigma_2,b}_{n^\prime}]\\
			=-if^{abc}
			\frac{(\frac{\sigma_1-k}{2}-n+\frac{\sigma_2-l}{2}-n^\prime)!
				(\frac{\sigma_1-k}{2}+n+\frac{\sigma_2-l}{2}+n^\prime)!}
			{(\frac{\sigma_1-k}{2}-n)!(\frac{\sigma_1-k}{2}+n)!(\frac{\sigma_2-l}{2}-n^\prime)! (\frac{\sigma_2-l}{2}+n^\prime)! } R^{k+l-1,\sigma_1+\sigma_2-1,c}_{n+n^\prime}.
		\end{gathered}
	\end{equation}	
	This is the simplest higher spin extension of the holographic symmetry algebra for gluon. 
	
	To relate it to the $S$-algebra we define the light transformed operator \cite{Strominger:2021mtt}
	\begin{equation}
		\label{eq:hs5}
		\begin{split}
			& S^{q,\sigma,a}_{m}=(q-m-1)!(q+m-1)!R^{-2q+\sigma+2,\sigma,a}_{m}.
		\end{split}
	\end{equation}	
	in terms of which the algebra \eqref{eq:hs4} becomes,
	\begin{equation}
		\label{eq:hs6}
		\begin{split}
			& [S^{p,\sigma_1,a}_{n},S^{q,\sigma_2,b}_{n^\prime}]=-if^{abc}S^{p+q-1,\sigma_1+\sigma_2-1,c}_{n+n^\prime}
		\end{split}
	\end{equation}
	where $p=1,\frac{3}{2},2,...$ and $1-p \le n \le p-1$ . We now discuss some of the infinite dimensional subalgebras of the higher spin $S$-algebra. 
	
	\section{Two copies of $S$-algebra}
	\subsubsection{$S$ algebra from soft spin-$1$ partcle}  
	
	The most obvious infinite dimensional subalgebra is the $S$-algebra generated by the spin-$1$ soft gluon \cite{Strominger:2021mtt} given by
	\begin{equation}
		\label{eq:sub1}
		\begin{split}
			&[S^{p,1,a}_{m},S^{q,1,b}_{n}]=-if^{abc}S^{p+q-1,1,c}_{m+n}
		\end{split}
	\end{equation}
	$S$ algebra generators act on $\{S^{p,\sigma,a}_{m}\}, \sigma>1$ as  
	\begin{equation}
		\label{eq:sub2}
		[S^{p,1,a}_{m},S^{q,\beta,b}_{n}]=-if^{abc}S^{p+q-1,\beta,c}_{m+n}
	\end{equation}
	So they transform in the adjoint representation of the $S$-algebra. 
	
	\subsubsection{$\tilde{S}$ algebra from soft higher spin particles} 
	The higher spin $S$-algebra \eqref{eq:hs6} has another infinite dimensional subalgebra which we call the $\tilde S$-algebra. $\tilde S$-algebra is isomorphic to the $S$-algebra and is generated by $\tilde{S}^{p,a}_{m}=S^{p,2p-1,a}_{m}$  which satisfy the algebra
	\begin{equation}
		\label{eq:sub3}
		[\tilde{S}^{p,a}_{m},\tilde{S}^{q,b}_{n}]=-if^{abc}\tilde{S}^{p+q-1,c}_{m+n}, \ p=1,\frac{3}{2},\cdots
	\end{equation} 
	\section{Leading OPE from Higher Spin Yang-Mills (HSYM) amplitude}
	
	One of the most interesting questions is whether there is an interacting higher spin theory in the four dimensional asymptotically flat space time whose asymptotic (soft) symmetry algebra is given by \eqref{eq:hs6} (or \eqref{hsa}). It is likely that no such theory exists if we want the theory to be local in the bulk. So we have to consider non-local theories. However, this does not mean that holographic dual theory will also be non-local. This is supported by the fact that the Vasiliev higher spin gauge theory on AdS, although non-local, has local holographic dual. 
	
	In \cite{Adamo:2022lah} the authors constructed an interesting example of an interacting theory of colored higher spin particles which they called the Higher Spin Yang-Mills (HSYM) theory. In this paper they also proposed a formula for the tree level MHV amplitudes with two negative helicity colored particles with arbitrary spin and any number of positive helicity colored particles of arbitrary spin. They have argued that this scattering formula cannot be reproduced by any local bulk theory. In this section we show that the leading celestial OPE between two outgoing positive helicity particles of arbitrary spin computed from the ($4$-point) MHV formula proposed in \cite{Adamo:2022lah} matches exactly with \eqref{eq:hs2}. The subleading terms of $\mathcal{O}(\frac{\bar z^n}{z})$ are completely determined by the $SL(2,\mathbb{R})_R$ invariance once the leading term is known. This shows that the MHV formula for arbitrary helicity proposed in \cite{Adamo:2022lah} has indeed the extended symmetry \eqref{eq:hs6}. Let us now describe the calculation.

	The momentum space color-dressed 4-point tree level MHV amplitude in Higher-Spin Yang-Mills (HSYM) theory is given by the following expression,
	\begin{equation}
		\label{eq:4p mspace amp1}
		\begin{gathered}
			\mathcal{A}_{4}(1^{-,a_1}_{\sigma_1}2^{-,a_2}_{\sigma_2}3^{+,a_3}_{\sigma_3}4^{+,a_4}_{\sigma_4})\\
			=g^2
			\left(f^{a_1a_2x}f^{xa_3a_4} A_{4}(1^{-}_{\sigma_1}2^{-}_{\sigma_2}3^{+}_{\sigma_3}4^{+}_{\sigma_4}) +
			f^{a_1a_3x}f^{xa_2a_4} 
			A_{4}(1^{-}_{\sigma_1}3^{+}_{\sigma_3}2^{-}_{\sigma_2}4^{+}_{\sigma_4})
			\right) \delta\left(\sum_{i=1}^{4}p^{\mu}_{i}\right)\\
		\end{gathered}		
	\end{equation} 
	where $f^{abc}$ are the structure constants of the gauge group and $p^{\mu}_{i}$ is the four momentum of the $i$-th particle. The color-stripped partial amplitude, $A_{4}(1^{-}_{\sigma_1}2^{-}_{\sigma_2}3^{+}_{\sigma_3}4^{+}_{\sigma_4})$ is given by  \cite{Adamo:2022lah},
	\begin{equation}
		\label{eq:amp1234}
		\begin{gathered}
			A_{4}(1^{-}_{\sigma_1}2^{-}_{\sigma_2}3^{+}_{\sigma_3}4^{+}_{\sigma_4})\\
			=\frac{\langle 12 \rangle^4}{\langle 12 \rangle \langle 23 \rangle \langle 34 \rangle \langle 41 \rangle}
			\left[
			\delta_{0,-\sigma_1+\sigma_2+\sigma_3+\sigma_4-2} 
			\left( \frac{\langle 12 \rangle^{\sigma_2+\sigma_3+\sigma_4-3}}
			{\langle 23 \rangle^{\sigma_3-1} 
				\langle 24 \rangle^{\sigma_4-1}}\right)^2 +
			\delta_{0,\sigma_1-\sigma_2+\sigma_3+\sigma_4-2} 
			\left( \frac{\langle 12 \rangle^{\sigma_1+\sigma_3+\sigma_4-3}}
			{\langle 13 \rangle^{\sigma_3-1} 
				\langle 14 \rangle^{\sigma_4-1}}\right)^2              
			\right] 
		\end{gathered}
	\end{equation}
	
	and the amplitude $A_{4}(1^{-}_{\sigma_1}3^{+}_{\sigma_3}2^{-}_{\sigma_2}4^{+}_{\sigma_4})$ can be written as (see Appendix \ref{app:check1} for the proof),
	\begin{equation}
		\label{eq:bcj applied}
		A_{4}(1^{-}_{\sigma_1}3^{+}_{\sigma_3}2^{-}_{\sigma_2}4^{+}_{\sigma_4})
		=\frac{s_{12}}{s_{13}}A_{4}(1^{-}_{\sigma_1}2^{-}_{\sigma_2}3^{+}_{\sigma_3}4^{+}_{\sigma_4})
	\end{equation}
	where, $s_{ij}=(p_i+p_j)^2$. We parameterize massless momenta and spinor variables in split signature $(-,+,-,+)$ as,
	\begin{equation}
		\label{eq:momentum param}
		p^{\mu}_{i}=\epsilon_i\omega_i q(z_i,\bar{z}_i)=\epsilon_i\omega_i
		(1+z_i\bar{z}_{i},z_{i}+\bar{z}_{i},z_{i}-\bar{z}_{i},1-z_i\bar{z}_{i}),
	\end{equation} 
	\begin{equation}
		\label{eq: spinor variables}
		\langle ij \rangle=-2 \epsilon_{i} \epsilon_{j}\sqrt{\omega_i \omega_j}z_{ij},~~
		[ij]=2 \sqrt{\omega_i \omega_j}\bar{z}_{ij}.
	\end{equation}
	where $\omega_i \in [0,\infty)$ represents the energy of the $i$-th massless particle and $\epsilon_{i}=+1(-1)$ corresponds to an outgoing (incoming) particle. In split signature, $z_i$ and $\bar{z}_{i}$ are two independent real variables.
    We will consider $(1,2)$ to be incoming and $(3,4)$ to be outgoing and perform further calculations.
	Using momentum conservation, one can write \eqref{eq:bcj applied} as,
	\begin{equation}
		\label{eq:bcj applied2}
		A_{4}(1^{-}_{\sigma_1}3^{+}_{\sigma_3}2^{-}_{\sigma_2}4^{+}_{\sigma_4})
		=-\frac{z_{12}z_{34}}{z_{13}z_{24}}A_{4}(1^{-}_{\sigma_1}2^{-}_{\sigma_2}3^{+}_{\sigma_3}4^{+}_{\sigma_4}).
	\end{equation}
	Substituting \eqref{eq:bcj applied2} in \eqref{eq:4p mspace amp1} and expressing the amplitude in $(\omega_i,z_i,\bar{z}_i)$ variables, we get
	\begin{equation}
		\label{eq:4 mspace amp2}
		\begin{gathered}
			\mathcal{A}_{4}(1^{-,a_1}_{\sigma_1}2^{-,a_2}_{\sigma_2}3^{+,a_3}_{\sigma_3}4^{+,a_4}_{\sigma_4})\\
			=\ g^2 \frac{\langle 12 \rangle^4}{\langle 12 \rangle 
				\langle 23 \rangle
				\langle 34 \rangle
				\langle 41 \rangle }
			\left(f^{a_1a_2x}f^{xa_3a_4}- \frac{z_{12}z_{34}}{z_{13}z_{24}}f^{a_1a_3x}f^{xa_2a_4} \right)\\
			 \times
			 \left[
			\delta_{0,-\sigma_1+\sigma_2+\sigma_3+\sigma_4-2} 
			\left( \frac{\langle 12 \rangle^{\sigma_2+\sigma_3+\sigma_4-3}}
			{\langle 23 \rangle^{\sigma_3-1} 
				\langle 24 \rangle^{\sigma_4-1}}\right)^2 +
			\delta_{0,\sigma_1-\sigma_2+\sigma_3+\sigma_4-2} 
			\left( \frac{\langle 12 \rangle^{\sigma_1+\sigma_3+\sigma_4-3}}
			{\langle 13 \rangle^{\sigma_3-1} 
				\langle 14 \rangle^{\sigma_4-1}}\right)^2              
			\right] \\
			 \times
			\delta^{(4)}\left( 
			\omega_1 q(z_1,\bar{z}_1)+
			\omega_2 q(z_2,\bar{z}_2)-
			\omega_3 q(z_3,\bar{z}_3)-
			\omega_4 q(z_4,\bar{z}_4)
			\right)\\
		=-g^2 \left(f^{a_1a_2x}f^{xa_3a_4}- \frac{z_{12}z_{34}}{z_{13}z_{24}}f^{a_1a_3x}f^{xa_2a_4} \right)\\
		\times
		 \Bigg(
	(-1)^{2 (\sigma_1+\sigma_3+ \sigma_4)} 4^{\sigma_1-1} \omega _1^{\sigma_1} \omega _2^{\sigma_1+\sigma_3+\sigma_4-2} \omega _3^{-\sigma_3} \omega _4^{-\sigma_4}\frac{z_{12}^{2(\sigma_1+\sigma_3+\sigma_4)-3}}{z_{13}^{2\sigma_3-2}z_{14}^{2\sigma_4-1}z_{23}z_{34}}\delta _{0,\sigma_1-\sigma_2+\sigma_3+\sigma_4-2}
	\\
	+(-1)^{2 (\sigma_2+ \sigma_3+ \sigma_4)} 4^{\sigma_2-1} \omega _1^{\sigma_2+\sigma_3+\sigma_4-2} \omega _2^{\sigma_2} \omega _3^{-\sigma_3} \omega _4^{-\sigma_4}\frac{z_{12}^{2(\sigma_2+\sigma_3+\sigma_4)-3}}{z_{23}^{2\sigma_3-1}z_{24}^{2\sigma_4-2}z_{14}z_{34}}\delta _{0,-\sigma_1+\sigma_2+\sigma_3+\sigma_4-2}\Bigg) \\
	\times
		\delta^{(4)}\left( 
		\omega_1 q(z_1,\bar{z}_1)+
		\omega_2 q(z_2,\bar{z}_2)-
		\omega_3 q(z_3,\bar{z}_3)-
		\omega_4 q(z_4,\bar{z}_4)
		\right)
\end{gathered}
	\end{equation}
	
	The momentum conserving delta function in the above expression can be represented as,
	\begin{equation}
		\begin{gathered}
		\delta^{(4)}\left( 
		\omega_1 q(z_1,\bar{z}_1)+
		\omega_2 q(z_2,\bar{z}_2)-
		\omega_3 q(z_3,\bar{z}_3)-
		\omega_4 q(z_4,\bar{z}_4)
		\right)\\
		=
		\frac{1}{4 \omega^{\star}_1 \omega^{\star}_2 z^{2}_{12}}                               
		\delta\left(\omega_1-\omega^{\star}_1\right)
		\delta\left(\omega_2-\omega^{\star}_2\right)
		\delta(\bar{z}_{14}+\frac{\omega_3}{\omega^{\star}_1} \frac{z_{23}}{z_{12}}\bar{z}_{34})
		\delta(\bar{z}_{24}-\frac{\omega_3}{\omega^{\star}_2} \frac{z_{13}}{z_{12}}\bar{z}_{34})
		\end{gathered}
	\end{equation} 
where		
	\begin{equation}
		\label{eq:omega1}
		\omega^{\star}_{1}=-\frac{z_{24}}{z_{12}}(\omega_3+\omega_4)+\frac{z_{34}}{z_{12}}\omega_3
	\end{equation}
	\begin{equation}
		\label{eq:omega2}
		\omega^{\star}_{2}=\frac{z_{14}}{z_{12}}(\omega_3+\omega_4)-\frac{z_{34}}{z_{12}}\omega_3.
	\end{equation}
	The corresponding celestial amplitude is obtained by performing Mellin transforms over the energies of external states in the amplitude \eqref{eq:4 mspace amp2},
	\begin{equation}
		\label{eq:mellin amp}
		\begin{gathered}
		\widetilde{\mathcal{M}}_{4}
		 (1^{\sigma_1,a_1}_{\Delta_1,-},2^{\sigma_2,a_2}_{\Delta_2,-},
		  3^{\sigma_3,a_3}_{\Delta_3,+},4^{\sigma_4,a_4}_{\Delta_4,+})\\
		 =\langle \mathcal{O}^{\sigma_1,a_1}_{\Delta_1,-}(z_1,\bar{z}_1)
		  \mathcal{O}^{\sigma_2,a_2}_{\Delta_2,-}(z_2,\bar{z}_2)
		  \mathcal{O}^{\sigma_3,a_3}_{\Delta_3,+}(z_3,\bar{z}_3)
		  \mathcal{O}^{\sigma_4,a_4}_{\Delta_4,+}(z_4,\bar{z}_4) \rangle\\
		=\int_{0}^{\infty} \prod_{i=1}^{4}d\omega_i \omega^{\Delta_i-1}_{i}
		  \mathcal{A}_{4}(1^{-,a_1}_{\sigma_1}2^{-,a_2}_{\sigma_2}3^{+,a_3}_{\sigma_3}4^{+,a_4}_{\sigma_4}).\\
		\end{gathered}
	\end{equation}

	To obtain the OPE between $\mathcal{O}^{\sigma_3,a_3}_{\Delta_3,+}(z_3,\bar{z}_3)$ and $\mathcal{O}^{\sigma_4,a_4}_{\Delta_4,+}(z_4,\bar{z}_4)$, we expand the above Mellin amplitude \eqref{eq:mellin amp} in the OPE limit $z_3 \rightarrow z_4$ and $\bar{z}_3 \rightarrow \bar{z}_{4}$. The leading order term of \eqref{eq:mellin amp} in the OPE expansion is given by,
	\begin{equation}
		\label{eq:lead1}
		\begin{gathered}
			\hspace{-2cm}
			\widetilde{\mathcal{M}}_{4}(1^{\sigma_1,a_1}_{\Delta_1,-},2^{\sigma_2,a_2}_{\Delta_2,-},3^{\sigma_3,a_3}_{\Delta_3,+},4^{\sigma_4,a_4}_{\Delta_4,+})\\
			=-\frac{g^2 \pi }{z_{34}} f^{a_1 a_2 x}f^{x a_3 a_4} 
			 B\left(\Delta _3-\sigma_3,\Delta _4-\sigma_4\right)
			\delta \left(\bar{z}_{14}\right) \delta \left(\bar{z}_{24}\right)\theta \left(\frac{z_{14}}{z_{12}}\right) \theta \left(\frac{z_{42}}{z_{12}}\right) \\
			 \hspace{-1cm}
			 \times 
			 \Bigg[
			 2^{2 \sigma_1-3} (-1)^{\Delta _1+3 \sigma_1+2 \left(\sigma_3+\sigma_4\right)} 
			 \delta _{0,\sigma_1-\sigma_2+\sigma_3+\sigma_4-2}
			\frac{z_{12}^{\sigma_3+\sigma_4-\Delta_1-\Delta_2+1}}{z_{14}^{\sigma_3+\sigma_4-\sigma_1-\Delta_2+1}z_{24}^{3-\Delta_1-\sigma_1}}
			\delta \left(\sum_{i=1}^{4}\Delta_i+2 \sigma_1-6\right)\\
			+2^{2 \sigma_2-3} (-1)^{\Delta _1+3 \sigma_2+3 \sigma_3+3 \sigma_4}
			\delta _{0,-\sigma_1+\sigma_2+\sigma_3+\sigma_4-2}
			 \frac{z_{12}^{\sigma_3+\sigma_4-\Delta_1-\Delta_2+1}}{z_{14}^{3-\Delta_2-\sigma_2}z_{24}^{\sigma_3+\sigma_4-\sigma_2-\Delta_1+1}}			 
			 \delta \left(\sum_{i=1}^{4}\Delta_i+2 \sigma_2-6\right)
			\Bigg]\\
			+\cdots 
		\end{gathered}
	\end{equation} 
	The tree-level three point celestial MHV amplitude in HSYM theory is given by, 
	\begin{equation}
		\label{eq:3p1}
		\begin{gathered}
			\hspace{-2cm}
			\widetilde{\mathcal{M}}_{3}(1^{\sigma_1,a_1}_{\Delta_1,-},2^{\sigma_2,a_2}_{\Delta_2,-},4^{\sigma_4,x}_{\Delta_4,+})\\
			\hspace{-1cm}
			=-ig \pi f^{a_1 a_2 x}  	
			\delta \left(\bar{z}_{14}\right) \delta \left(\bar{z}_{24}\right)
			\theta \left(\frac{z_{14}}{z_{12}}\right) \theta \left(\frac{z_{42}}{z_{12}}\right)  \\
			\times\Bigg[2^{2 \sigma_1-3} (-1)^{\Delta _1+3 \sigma_1+2 \sigma_4}
			\delta _{0,-\sigma_1+\sigma_2-\sigma_4+1} 		
			\frac{z_{12}^{\sigma_4-\Delta_1-\Delta_2+2}}{z_{14}^{\sigma_4-\sigma_1-\Delta_2+2}z_{24}^{3-\Delta_1-\sigma_1}}		
			\delta \left(\sum_{i\neq3}^{4}\Delta_i+2 \sigma_1-5\right)	\\
			-2^{2 \sigma_2-3} (-1)^{\Delta _1+3 \left(\sigma_2+\sigma_4\right)}
		\delta _{0,\sigma_1-\sigma_2-\sigma_4+1}
		\frac{z_{12}^{\sigma_4-\Delta_1-\Delta_2+2}}{z_{14}^{3-\Delta_2-\sigma_2}z_{24}^{\sigma_4-\sigma_2-\Delta_1+2}}			
		\delta \left(\sum_{i\neq3}^{4}\Delta_i+2 \sigma_2-5\right)
			\Bigg]
		\end{gathered}
	\end{equation}
	Now with following shifts in scaling dimension $\Delta_4$ and spin $\sigma_4$,
	\begin{equation}
		\label{eq:shifts1}
		\begin{split}
			& \Delta _4\to \Delta _4+\Delta _3-1 \\
			& \sigma_4\to \sigma_4+\sigma_3-1
		\end{split}
	\end{equation} 
	the 3-point amplitude \eqref{eq:3p1} becomes,
	\begin{equation}
		\label{eq:shifts2}
		\begin{gathered}
			\hspace{-2cm}
			\widetilde{\mathcal{M}}_{3}(1^{\sigma_1,a_1}_{\Delta_1,-},2^{\sigma_2,a_2}_{\Delta_2,-},4^{\sigma_3+\sigma_4-1,x}_{\Delta _3+\Delta _4-1,+})\\
			\hspace{-2cm}
			=-ig \pi f^{a_1 a_2 x}  	
			\delta \left(\bar{z}_{14}\right) \delta \left(\bar{z}_{24}\right)
			\theta \left(\frac{z_{14}}{z_{12}}\right) \theta \left(\frac{z_{42}}{z_{12}}\right) \\
			 \hspace{-1cm}
			 \times
			 \Bigg[2^{2 \sigma_1-3} (-1)^{\Delta _1+3 \sigma_1+2 \left(\sigma_3+\sigma_4\right)}
			\delta _{0,-\sigma_1+\sigma_2-\sigma_3-\sigma_4+2}
			\frac{z_{12}^{\sigma_3+\sigma_4-\Delta_1-\Delta_2+1}}{z_{14}^{\sigma_3+\sigma_4-\sigma_1-\Delta_2+1}z_{24}^{3-\Delta_1-\sigma_1}}		
			\delta \left(\sum_{i=1}^{4}\Delta_i+2 \sigma_1-6\right)\\
			+
			2^{2 \sigma_2-3} (-1)^{\Delta _1+3 \left(\sigma_2+\sigma_3+\sigma_4\right)}
			\delta _{0,\sigma_1-\sigma_2-\sigma_3-\sigma_4+2}
			\frac{z_{12}^{\sigma_3+\sigma_4-\Delta_1-\Delta_2+1}}{z_{14}^{3-\Delta_2-\sigma_2}z_{24}^{\sigma_3+\sigma_4-\sigma_2-\Delta_1+1}}			
			\delta \left(\sum_{i=1}^{4}\Delta_i+2 \sigma_2-6\right)
			\Bigg].
		\end{gathered}
	\end{equation}
	From \eqref{eq:lead1} and \eqref{eq:shifts2}, at the level of correlation functions we can write
	\begin{equation}
		\label{eq:correllator1}
		\begin{gathered}
			 \langle \mathcal{O}^{\sigma_1,a_1}_{\Delta_1,-}(z_1,\bar{z}_1)
			\mathcal{O}^{\sigma_2,a_2}_{\Delta_2,-}(z_2,\bar{z}_2)
			\mathcal{O}^{\sigma_3,a_3}_{\Delta_3,+}(z_3,\bar{z}_3)
			\mathcal{O}^{\sigma_4,a_4}_{\Delta_4,+}(z_4,\bar{z}_4) \rangle \\
			=-\frac{ig f^{a_3a_4 x}}{z_{34}}
			B\left(\Delta _3-\sigma_3,\Delta _4-\sigma_4\right)
			\langle \mathcal{O}^{\sigma_1,a_1}_{\Delta_1,-}(z_1,\bar{z}_1)
			\mathcal{O}^{\sigma_2,a_2}_{\Delta_2,-}(z_2,\bar{z}_2)
			\mathcal{O}^{\sigma_3+\sigma_4-1,x}_{\Delta_3+\Delta_4-1,+}(z_4,\bar{z}_4) \rangle +\cdots         
		\end{gathered}
	\end{equation}
	Hence, the leading order term in the OPE is given by,
	\begin{equation}
		\label{eq:lead_ope1}
		\mathcal{O}^{\sigma_3,a_3}_{\Delta_3,+}(z_3,\bar{z}_3)
		\mathcal{O}^{\sigma_4,a_4}_{\Delta_4,+}(z_4,\bar{z}_4)\sim
		-\frac{ig f^{a_3a_4 x}}{z_{34}} 
		B\left(\Delta _3-\sigma_3,\Delta _4-\sigma_4\right)
		\mathcal{O}^{\sigma_3+\sigma_4-1,x}_{\Delta_3+\Delta_4-1,+}(z_4,\bar{z}_4)
	\end{equation}
	which matches with the leading order term in \eqref{eq:hs2}. What is important here is the helicity selection rule $O^{\sigma_1}\times O^{\sigma_2}\sim O^{\sigma_1+\sigma_2 -1}$ which we can see is clearly reproduced here. 
	
		\section{Higher spin extension in curved background}
	In this section, we extend the analysis of higher-spin algebra \eqref{hsa} to curved backgrounds and compute the deformed algebra in the presence of the cosmological constant $\Lambda$.
	In \cite{Taylor:2023ajd}, the authors computed the positive helicity graviton-graviton OPE \eqref{eq:GG ope} in the presence of a nonzero cosmological constant which takes the form
	\begin{equation}
		\label{eq:L1}
		\begin{gathered}
			G^{+}_{\Delta_1}(z_1,\bar{z}_{1})G^{+}_{\Delta_2}(z_2,\bar{z}_{2})\\
		\sim 
		- \frac{\kappa}{2}\frac{1}{z_{12}} \sum_{n=0}^{\infty} B(\D_1 -1 + n, \D_2 - 1) \frac{\bar z_{12}^{n+1}}{n!} \bar\partial_2^{n} 
		G^{+}_{\D_1 +\D_2}(z_2,\bar z_2)\\
		+\frac{\kappa \Lambda}{2}\frac{1}{z^{2}_{12}} \sum_{n=0}^{\infty} B(\D_1-2+n,\D_2-2)\frac{\bar{z}^{n}_{12}}{n!} \bar{\partial}^{n}_{2}
		\left[
		(\D_1+\D_2)G^{+}_{\D_1+\D_2-2}(z_2,\bar{z}_{2})+
		\D_1 z_{12} \partial_{2} G^{+}_{\D_1+\D_2-2}(z_2,\bar{z}_{2})
		\right] 
		\end{gathered}
	\end{equation}	
	where $G^{+}_{\D}(z,\bar{z})$ denotes a positive helicity graviton primary operator of scaling dimension $\D$. 
	The OPE \eqref{eq:L1} displays only the singular terms of the $G^{+}G^{+}$ OPE in the limit $z_1\rightarrow z_2$. These terms completely determine the associated soft symmetry algebra.
	
	 Let us now analyze the singular terms appearing in the OPE in detail. The first term on the RHS of \eqref{eq:L1} corresponds to the standard flat-space contribution. This part of the singular OPE receives contributions exclusively from the graviton primary $G^{+}_{\D_1 +\D_2}(z_2,\bar z_2)$ and its $SL(2,\mathbb{R})_{R}$ descendants. Once the leading OPE coefficient is known, all subleading OPE coefficients $B(\D_1-1+n,\D_2-1)$ at order $\mathcal{O}\left(\frac{\bar{z}^{n+1}_{12}}{z_{12}}\right)$ are completely determined by $SL(2,\mathbb{R})_{R}$ invariance alone, as mentioned in section \ref{sec:hsym}.

	 At order $\mathcal{O}(\Lambda)$, however, the OPE begins to receive additional singular contributions from the graviton primary $G^{+}_{\D_1+\D_2-2}(z_2,\bar{z}_{2})$ and its descendants under both $SL(2,\mathbb{R})_{L}$ and $SL(2,\mathbb{R})_{R}$. The terms which appear on the third line of \eqref{eq:L1} is the contribution of the conformal family of $G^+_{\D_1+\D_2-2}$ to the $G^+_{\D_1}G^{+}_{\D_2}$ OPE and can be written as 
	 \be\label{contr}
	 \begin{gathered}
	 G^{+}_{\Delta_1}(z_1,\bar{z}_{1})G^{+}_{\Delta_2}(z_2,\bar{z}_{2}) \supset \\
	\frac{\kappa\Lambda}{2} \(\sum_{n=0}^{\infty} B(\D_1-2+n,\D_2-2)\frac{\bar{z}^{n}_{12}}{n!} \bar{\partial}^{n}_{2}\)
		\(
		(\D_1+\D_2)\frac{1}{z_{12}^2}+
		\D_1 \frac{1}{z_{12}} \partial_{2} + \mathcal{O}(z_{12}^0)\) G^{+}_{\D_1+\D_2-2}(z_2,\bar{z}_{2})
	 \end{gathered}
	  \ee
 The set of antiholomorphic terms in \eqref{contr} is completely fixed in terms of the leading OPE coefficient which is $B(\D_1-2,\D_2-2)$ by the $SL(2,\mathbb{R})_{R}$ invariance.  Similarly the set of holomorphic terms in \eqref{contr} is completely fixed in terms of the leading OPE coefficient $(\D_1+\D_2)$ by the $SL(2,\mathbb{R})_{L}$ invariance. To verify this let us write the holomorphic part of the OPE between two conformal primaries as 
       \begin{equation}
	 	\label{eq:phiphi}
	 	\begin{gathered}
	 		\phi_{h_i,\bar h_i}(z,\bar{z})\phi_{h_j,\bar h_j}(0,0)
	 		\supset \sum_{n=0}^{\infty}
	 		c^{n}_{ijk} z^{h_k-h_j-h_i+n} 
	 		({\partial^n}\phi_{h_k,\bar h_k})(0,0)
	 	\end{gathered}
	 \end{equation}
      Then it is well known that the $SL(2,\mathbb{R})_L$ invarinace leads to the following recursion relation      
	 \begin{equation}\label{rec}
	 	\begin{gathered}
	 		c^{n}_{ijk}
	 		=\frac{(h_k+h_i-h_j+n-1)\cdots(h_k+h_i-h_j+1)(h_k+h_i-h_j)}
	 		{n!(2h_k+n-1)(2h_k+n-2)\cdots (2h_k+1) 2h_k}c^{0}_{ijk}
	 	\end{gathered}
	 \end{equation} 
	 Now one can easily check that the holomorphic OPE coefficients $(\D_1+\D_2)$ and $\Delta_{1}$ appearing in \eqref{contr} can be identified with $c^{0}_{ijk}$ and $c^{1}_{ijk}$ and they satisfy the recursion relation \eqref{rec} with
	 \begin{equation}
	 	\begin{gathered}
	 		h_i=\frac{\D_1+2}{2},~
	 		h_j=\frac{\D_2+2}{2},~
	 		h_k=\frac{\D_1+\D_2}{2}
	 	\end{gathered}
	 \end{equation}
	 The same argument goes through for the antiholomorphic part of the OPE \eqref{contr}.


	 The above discussion allows us to rewrite the OPE \eqref{eq:L1} as
	\begin{equation}
		\label{eq:L2}
		\begin{gathered}
		 	G_{h_1,\bar h_1}(z_1,\bar z_1) G_{h_2,\bar h_2}(z_2,\bar z_2)\\
		 	 \sim - \frac{\kappa}{2} \frac{1}{z_{12}} \sum_{n=0}^{\infty} 
		 	 B(2\bar h_1 +1 + n, 2\bar h_2 +1) \frac{\bar z_{12}^{n+1}}{n!} \bar\partial_2^{n} G_{h_1+h_2-1, \bar h_1 +\bar h_2 +1}(z_2,\bar z_2)\\
		 	 +\frac{\kappa \Lambda}{2}\frac{1}{z^{2}_{12}} \sum_{n=0}^{\infty} B(2\bar h_1 + n, 2\bar h_2)\frac{\bar{z}^{n}_{12}}{n!} \bar{\partial}^{n}_{2}
		 	 \left[
		 	 (2h_1+2h_2-4)+
		 	 (2h_1-2) z_{12} \partial_{2}
		 	 \right]G_{h_1+h_2-2, \bar h_1 +\bar h_2 }(z_2,\bar z_2) 
		\end{gathered}
	\end{equation}
		In this form we can analytically continue \eqref{eq:L2} in $(h,\bar{h})$ to obtain a higher spin OPE in direct analogy with the flat-space construction. The singular terms in the celestial OPE between two higher spin particles now takes the following form, 
	\begin{equation}
		\label{eq:LGG OPE 1}
		\begin{gathered}
			G^{\sigma_1}_{\Delta_1}(z_1,\bar{z}_1)
			G^{\sigma_2}_{\Delta_2}(z_2,\bar{z}_2)\\
			\sim 
			- \frac{\kappa}{2}\frac{1}{z_{12}} \sum_{n=0}^{\infty} B(\D_1 -\sigma_1 +1 + n, \D_2 -\sigma_2 +1) \frac{\bar z_{12}^{n+1}}{n!} \bar\partial_2^{n} G^{\sigma_1 +\sigma_2 -2}_{\D_1 +\D_2}(z_2,\bar z_2)\\
			+\frac{\kappa \Lambda}{2} \frac{1}{z^{2}_{12}} 
			\sum_{n=0}^{\infty} B(\Delta_1-\sigma_1+n,\Delta_2-\sigma_2)
			\frac{\bar{z}^{n}_{12}}{n!} \bar{\partial}^{n}_{2}
			[ 
			(\Delta_1+\Delta_2+\sigma_1+\sigma_2-4)
			G^{\sigma_1+\sigma_2-2}_{\Delta_1+\Delta_2-2}(z_2,\bar{z}_2)\\
			+z_{12}(\Delta_{1}+\sigma_1-2)\partial_2
			G^{\sigma_1+\sigma_2-2}_{\Delta_1+\Delta_2-2}(z_2,\bar{z}_2)
		    ]
		\end{gathered}
	\end{equation}
	where $G^{\sigma}_{\Delta}$ denotes a higher spin celestial primary operator of dimension $\Delta$ and helicity $\sigma$. We take $\sigma$ to be a positive integer, with the restriction $\sigma \ge 2$ because the spin selection rule $G^{\sigma_1}\times G^{\sigma_2}\sim G^{\sigma_1+\sigma_2-2}$ remains unchanged after turning on the cosmological constant.
	
	Now to obtain the higher spin soft symmetry algebra 
    we first write down the OPE between two higher spin soft operators, defined through \eqref{eq:soft graviton definition}, by taking the soft limits in \eqref{eq:LGG OPE 1},
	\begin{equation}
		\label{eq:deformed ope}
		\begin{gathered}
			H^{k,\sigma_1}(z_1,\bar{z}_1) H^{l,\sigma_2}(z_2,\bar{z}_2)\\
			\sim 
			-\frac{\kappa}{2} \frac{1}{z_{12}} 
			\sum_{m=0}^{\sigma_1-k-1} 
			\frac{(\sigma_1+\sigma_2-k-l-m-2)!}{(\sigma_1-k-m-1)!(\sigma_2-1-l)!}
			\frac{\bar{z}^{m+1}_{12}}{m!}\bar{\partial}^{m}_{2}H^{k+l,\sigma_1+\sigma_2-2}(z_2,\bar{z}_{2})\\
			+ \frac{\kappa \Lambda}{2} \frac{1}{z^{2}_{12}}
			\sum_{m=0}^{\sigma_1-k} 
			\frac{(\sigma_1+\sigma_2-k-l-m)!}{(\sigma_1-k-m)!(\sigma_2-l)!}
			\frac{\bar{z}^{m}_{12}}{m!}
			\bar{\partial}^{m}_{2}\\
			\times 
			\left[(k+l+\sigma_1+\sigma_2-4)
			+(k+\sigma_1-2)z_{12}\partial_2
			\right]
			H^{k+l-2,\sigma_1+\sigma_2-2}(z_2,\bar{z}_{2})
		\end{gathered}
	\end{equation}
	Substituting the mode expansion \eqref{eq:mode expansion} in \eqref{eq:deformed ope}, one obtains the OPE between two currents,
	\begin{equation}
		\label{current current ope}
		\begin{gathered}
			H^{k,\sigma_1}_{m}(z_1) H^{l,\sigma_2}_{n}(z_2)
		     \sim 
			-\frac{\kappa}{2} \frac{1}{z_{12}} [n(\sigma_1-k)-m(\sigma_2-l)] \\
			\times     
			\frac{(\frac{\sigma_1-k}{2}+\frac{\sigma_2-l}{2}-m-n-1)!(\frac{\sigma_1-k}{2}+\frac{\sigma_2-l}{2}+m+n-1)!}
			{(\frac{\sigma _1-k}{2}+m)!(\frac{\sigma _1-k}{2}-m)!(\frac{\sigma _2-l}{2}+n)!(\frac{\sigma _2-l}{2}-n)!}H^{k+l,\sigma_1+\sigma_2-2}_{m+n}(z_2)  
			\\
			+\frac{\kappa \Lambda}{2} \frac{1}{z^{2}_{12}}
			\frac{(\frac{\sigma_1-k}{2}+\sigma_2-l+m)!
				  (\frac{\sigma_1-k}{2}+\frac{\sigma_2-l}{2}-m-n)!}
			{(\frac{\sigma_1-k}{2}+m)!
			 (\frac{\sigma_1-k}{2}-m)!
			 (\frac{\sigma_2-l}{2}-n)!
			 (\sigma_2-l)!}
			\\
			 \times _2F_1\left(-\frac{\sigma_1-k}{2}-m,-\frac{\sigma_2-l}{2}+n; -\frac{\sigma_1-k}{2}-\sigma_2+l-m;1\right)\\
			\times \Bigg[(k+l+\sigma_1+\sigma_2-4)+(k+\sigma _1-2)z_{12}\partial_2\Bigg]  H^{k+l-2,\sigma_1+\sigma_2-2}_{m+n}(z_2)    
		\end{gathered}
	\end{equation}
	where $_2F_1(a,b;c;1)$ is the Gaussian hypergeometric function, which is given by
	\begin{equation}
		\label{eq:hpg1}
		_2F_1(a,b;c;1)=\frac{\Gamma(c)\Gamma(c-a-b)}{\Gamma(c-a)\Gamma(c-b)}, ~~~~~\mathcal{R}(c)>\mathcal{R}(a+b).
	\end{equation}
	The currents $H^{k,\sigma}_{m}(z)$ admit further Laurent series expansion in $z$,
	\begin{equation}
		\label{eq:hmode1}
		H^{k,\sigma}_{m}(z)=
		\sum_{\alpha \in \mathbb{Z}-\frac{k+\sigma}{2}} 
		\frac{H^{k,\sigma}_{\alpha,m}}{z^{\alpha+\frac{k+\sigma}{2}}}
	\end{equation}
	where modes $H^{k,\sigma}_{\alpha,m}$ of the soft current are given by,
	\begin{equation}
		\label{eq:hmode2}
		H^{k,\sigma}_{\alpha,m}=
		\oint_{z=0} \frac{dz}{2 \pi i} \hspace{2mm}
		z^{\alpha+\frac{k+\sigma}{2}-1}H^{k,\sigma}_{m}(z).
	\end{equation}
	One can evaluate the commutator between these modes as,
	\begin{equation}
		\label{eq:commutator1}
		\begin{gathered}
		[H^{k,\sigma_1}_{a,m},H^{l,\sigma_2}_{b,n}]\\
		=
		\oint \frac{dz_{2}}{2 \pi i} \hspace{2mm}
		z^{b+\frac{k+\sigma_2}{2}-1}_{2}
		\oint \frac{dz_{1}}{2 \pi i} \hspace{2mm}
		z^{a+\frac{k+\sigma_1}{2}-1}_{1}
		H^{k,\sigma_1}_{m}(z_1)	H^{l,\sigma_2}_{n}(z_2).
		\end{gathered}
	\end{equation}
	The algebra of these modes is given by, 
	\begin{equation}
		\label{eq:Lw1}
		\begin{gathered}
			[H^{k,\sigma_1}_{a,m},H^{l,\sigma_2}_{b,n}]\\
			=-\frac{\kappa}{2}[n(\sigma_1-k)-m(\sigma_2-l)]     
			\frac{(\frac{\sigma_1-k}{2}+\frac{\sigma_2-l}{2}-m-n-1)!(\frac{\sigma_1-k}{2}+\frac{\sigma_2-l}{2}+m+n-1)!}
			{(\frac{\sigma _1-k}{2}+m)!(\frac{\sigma _1-k}{2}-m)!(\frac{\sigma _2-l}{2}+n)!(\frac{\sigma _2-l}{2}-n)!}H^{k+l,\sigma_1+\sigma_2-2}_{a+b,m+n}\\
			+\frac{\kappa \Lambda}{2} [a \left(l+\sigma _2-2\right)-b \left(k+\sigma _1-2\right)]
			\frac{(\frac{\sigma_1-k}{2}+\frac{\sigma_2-l}{2}-m-n)!(\frac{\sigma_1-k}{2}+\frac{\sigma_2-l}{2}+m+n)!}
			{(\frac{\sigma _1-k}{2}+m)!(\frac{\sigma _1-k}{2}-m)!(\frac{\sigma _2-l}{2}+n)!(\frac{\sigma _2-l}{2}-n)!}
			H^{k+l-2,\sigma_1+\sigma_2-2}_{a+b,m+n}.
		\end{gathered}
	\end{equation}
	We now define light transformed higher spin operators as \cite{Strominger:2021mtt},
	\begin{equation}
		\label{eq:redef1}
		w^{p,\sigma}_{a,m}=\frac{1}{\kappa} (p-1-m)!(p+m-1)!H^{-2p+2+\sigma,\sigma}_{a,m}.
	\end{equation}
	where $p = 1, \frac{3}{2}, 2,....$ and $1-p \le m \le p-1$.
	The algebra of these generators is given by,
	\begin{equation}
		\label{eq:wcos1}
		\begin{split}
			&[w^{p,\sigma_1}_{a,m},w^{q,\sigma_2}_{b,n}]=[m(q-1)-n(p-1)]w^{p+q-2,\sigma_1+\sigma_2-2}_{a+b,m+n}-\Lambda[a(q-\sigma_2)-b(p-\sigma_1)]w^{p+q-1,\sigma_1+\sigma_2-2}_{a+b,m+n}.
		\end{split}
	\end{equation}
	This is the higher spin analog of the deformed $w_{1+\infty}$ in curved background derived in \cite{Taylor:2023ajd}, which can be obtained by setting $\sigma_1=\sigma_2=2$ in \eqref{eq:wcos1}. In Appendix \ref{app:check Jacobi},
	we have shown explicitly that the generators $\{w^{p,\sigma}_{a,m}\}$ satisfy the Jacobi identity.

	\section*{Acknowledgment}
	We are grateful to Dmitry Ponomarev for helpful correspondence. The work of SB is supported by the  Swarnajayanti Fellowship (File No- SB/SJF/2021-22/14) of the Department of Science and Technology and ANRF, India. The work of RM is supported by ANRF-CRG project (CRG/2022/006165). SP is supported by the INSA Senior scientist position at NISER, Bhubaneswar, India through the Grant number INSA/SP/SS/2023.
	
       \appendix
	\section{Some checks}
	\label{app:check1}
	In this appendix we check the identity \eqref{eq:bcj applied}.
	The $n$-point color-stripped MHV amplitude in Higher-Spin Yang-Mills (HSYM) theory is given by  
	\cite{Adamo:2022lah},
	\begin{equation}
		\label{eq:npoint}
		\begin{gathered}
			A_{n}(1^{+}_{\sigma_1},\cdots,i^{-}_{\sigma_i},\cdots, j^{-}_{\sigma_j},\cdots,n^{+}_{\sigma_n})=
			\frac{\langle ij \rangle^4}{\langle 12 \rangle \langle 23 \rangle \cdots \langle n1 \rangle}\\
			\times 
			\left[
			\delta_{0,-\sigma_i+\sigma_j-n+2+\sum_{a\neq i,j} \sigma_a} 
			\left( \frac{\langle ij \rangle^{\sigma_j-n+1+\sum_{a\neq i,j}\sigma_a }}
			{\prod_{b\neq i,j}\langle jb \rangle^{\sigma_b-1}}\right)^2 +
			\delta_{0,\sigma_i-\sigma_j-n+2+\sum_{a\neq i,j}\sigma_a} 
			\left( \frac{\langle ij \rangle^{\sigma_i-n+1+\sum_{a\neq i,j}\sigma_a }}
			{\prod_{b\neq i,j}\langle ib \rangle^{\sigma_b-1}}\right)^2              
			\right] 
		\end{gathered}
	\end{equation}
	 The four-point color-stripped partial amplitude, $A_{4}(1^{-}_{\sigma_1}2^{-}_{\sigma_2}3^{+}_{\sigma_3}4^{+}_{\sigma_4})$ takes the following form,
	\begin{equation}
		\label{eq:ampp1234}
		\begin{gathered}
			A_{4}(1^{-}_{\sigma_1}2^{-}_{\sigma_2}3^{+}_{\sigma_3}4^{+}_{\sigma_4})\\
			=\frac{\langle 12 \rangle^4}{\langle 12 \rangle \langle 23 \rangle \langle 34 \rangle \langle 41 \rangle}
			\left[
			\delta_{0,-\sigma_1+\sigma_2+\sigma_3+\sigma_4-2} 
			\left( \frac{\langle 12 \rangle^{\sigma_2+\sigma_3+\sigma_4-3}}
			{\langle 23 \rangle^{\sigma_3-1} 
				\langle 24 \rangle^{\sigma_4-1}}\right)^2 +
			\delta_{0,\sigma_1-\sigma_2+\sigma_3+\sigma_4-2} 
			\left( \frac{\langle 12 \rangle^{\sigma_1+\sigma_3+\sigma_4-3}}
			{\langle 13 \rangle^{\sigma_3-1} 
				\langle 14 \rangle^{\sigma_4-1}}\right)^2              
			\right] 
		\end{gathered}	
	\end{equation}
	Now we compute the RHS of \eqref{eq:bcj applied},
	\begin{equation}
		\label{eq:rhs}
		\begin{gathered}
		\frac{s_{12}}{s_{13}}A_{4}(1^{-}_{\sigma_1}2^{-}_{\sigma_2}3^{+}_{\sigma_3}4^{+}_{\sigma_4})
		=\frac{\langle 12 \rangle[12]}{\langle 13 \rangle [13]}
		\frac{\langle 12 \rangle^4}{\langle 12 \rangle \langle 23 \rangle \langle 34 \rangle \langle 41 \rangle}\\
		\times 
		\left[
		\delta_{0,-\sigma_1+\sigma_2+\sigma_3+\sigma_4-2} 
		\left( \frac{\langle 12 \rangle^{\sigma_2+\sigma_3+\sigma_4-3}}
		{\langle 23 \rangle^{\sigma_3-1} 
			\langle 24 \rangle^{\sigma_4-1}}\right)^2 +
		\delta_{0,\sigma_1-\sigma_2+\sigma_3+\sigma_4-2} 
		\left( \frac{\langle 12 \rangle^{\sigma_1+\sigma_3+\sigma_4-3}}
		{\langle 13 \rangle^{\sigma_3-1} 
			\langle 14 \rangle^{\sigma_4-1}}\right)^2              
		\right] \\
		=\frac{\langle 42 \rangle[21]}{ \langle 43 \rangle[31]}
		\frac{\langle 12 \rangle^4}{\langle 13 \rangle \langle 23 \rangle \langle 24 \rangle \langle 41 \rangle}\\
		\times\left[
		\delta_{0,-\sigma_1+\sigma_2+\sigma_3+\sigma_4-2} 
		\left( \frac{\langle 12 \rangle^{\sigma_2+\sigma_3+\sigma_4-3}}
		{\langle 23 \rangle^{\sigma_3-1} 
			\langle 24 \rangle^{\sigma_4-1}}\right)^2 +
		\delta_{0,\sigma_1-\sigma_2+\sigma_3+\sigma_4-2} 
		\left( \frac{\langle 12 \rangle^{\sigma_1+\sigma_3+\sigma_4-3}}
		{\langle 13 \rangle^{\sigma_3-1} 
			\langle 14 \rangle^{\sigma_4-1}}\right)^2              
		\right] 
	    \end{gathered}
	\end{equation}
	where 
	\begin{equation}
		\label{eq:Mandelstam}
		s_{ij}=(p_i+p_j)^2=2 p_i\cdot p_j=\langle ij \rangle [ij].
	\end{equation}
     Using four momentum conservation equation 
     \begin{equation}
     	\label{eq:conservation1}
     	1\rangle[1 + 2\rangle[2+3\rangle[3+	4\rangle[4=0
     \end{equation}
     and contracting it with $\langle 4$ from left and with $1]$ from right, we get the following relation,
     \begin{equation}
     	\label{eq:conservation2}
     	\begin{gathered}
     	\langle 42 \rangle[21]=-\langle 43 \rangle[31]\\
     	\Rightarrow \frac{	\langle 42 \rangle[21]}{\langle 43 \rangle[31]}=-1.
     	\end{gathered}
     \end{equation}
     Substituting \eqref{eq:conservation2} in \eqref{eq:rhs}, we can write
     \begin{equation}
     	\label{eq:rhs2}
     	\begin{gathered}
     	 \frac{s_{12}}{s_{13}}A_{4}(1^{-}_{\sigma_1}2^{-}_{\sigma_2}3^{+}_{\sigma_3}4^{+}_{\sigma_4})=
     	 	\frac{\langle 12 \rangle^4}{\langle 13 \rangle \langle 32 \rangle \langle 24 \rangle \langle 41 \rangle}\\
     	 \times\left[
     	 \delta_{0,-\sigma_1+\sigma_2+\sigma_3+\sigma_4-2} 
     	 \left( \frac{\langle 12 \rangle^{\sigma_2+\sigma_3+\sigma_4-3}}
     	 {\langle 23 \rangle^{\sigma_3-1} 
     	 	\langle 24 \rangle^{\sigma_4-1}}\right)^2 +
     	 \delta_{0,\sigma_1-\sigma_2+\sigma_3+\sigma_4-2} 
     	 \left( \frac{\langle 12 \rangle^{\sigma_1+\sigma_3+\sigma_4-3}}
     	 {\langle 13 \rangle^{\sigma_3-1} 
     	 	\langle 14 \rangle^{\sigma_4-1}}\right)^2              
     	 \right] \\
     	 =A_{4}(1^{-}_{\sigma_1}3^{+}_{\sigma_3}2^{-}_{\sigma_2}4^{+}_{\sigma_4})
     	\end{gathered}
     \end{equation} 
     
\section{Jacobi Identity}
	\label{app:check Jacobi}
	In this appendix, we explicitly check that generators $\{w^{p,\sigma}_{a,m}\}$ of the higher spin soft symmetry algebra \eqref{eq:wcos1} in curved space, satisfy the Jacobi identity
	\begin{equation}
		\label{eq:jacobi1}
		[[w^{p,\sigma_1}_{a,m},w^{q,\sigma_2}_{b,n}],w^{r,\sigma_3}_{c,s}]+
		[[w^{q,\sigma_2}_{b,n},w^{r,\sigma_3}_{c,s}],w^{p,\sigma_1}_{a,m}]+
		[[w^{r,\sigma_3}_{c,s},w^{p,\sigma_1}_{a,m}],w^{q,\sigma_2}_{b,n}]=0.
	\end{equation}
	
	We use the algebra \eqref{eq:wcos1} to evaluate the following commutators that appear in the above equation,
	\begin{equation}
		\label{eq:term1}
		\begin{gathered}
			[[w^{p,\sigma_1}_{a,m},w^{q,\sigma_2}_{b,n}],w^{r,\sigma_3}_{c,s}]\\
			=(m (q-1)-n (p-1))
			[w^{p+q-2,\sigma _1+\sigma _2-2}_{a+b,m+n},w^{r,\sigma_3}_{c,s}]
			-\Lambda  \left(a \left(q-\sigma _2\right)-b \left(p-\sigma _1\right)\right)
			[w^{p+q-1,\sigma _1+\sigma _2-2}_{a+b,m+n},w^{r,\sigma_3}_{c,s}]
			\\
			=(m (q-1)-n (p-1))\Bigg( 
			((r-1) (m+n)-s (p+q-3))
			w^{p+q+r-4,\sigma _1+\sigma _2+\sigma _3-4}_{a + b + c, m + n + s} \\
			+\Lambda  \left(-r (a+b)+\sigma _3 (a+b)+c (p+q)-c \left(\sigma _1+\sigma _2\right)\right)
			w^{p+q+r-3,\sigma _1+\sigma _2+\sigma _3-4}_{a + b + c, m + n + s} 
			\Bigg)\\
			-\Lambda  \left(a (q-\sigma _2)-b(p- \sigma _1)\right)
			\Bigg(
			((r-1) (m+n)-s (p+q-2))
			w^{p+q+r-3,\sigma _1+\sigma _2+\sigma _3-4}_{a + b + c, m + n + s} \\
			+\Lambda  \left(-r (a+b)+\sigma _3 (a+b)+c (p+q+1)-c \left(\sigma _1+\sigma _2\right)\right)
			w^{p+q+r-2,\sigma _1+\sigma _2+\sigma _3-4}_{a + b + c, m + n + s}
			\Bigg),
		\end{gathered}
	\end{equation}
	\begin{equation}
		\label{eq:term2}
		\begin{gathered}
			[[w^{q,\sigma_2}_{b,n},w^{r,\sigma_3}_{c,s}],w^{p,\sigma_1}_{a,m}]\\
			=(n (r-1) -s(q-1))\Bigg(
			((p-1) (n+s)-m (q+r-3))
			w^{p+q+r-4,\sigma _1+\sigma _2+\sigma _3-4}_{a + b + c, m + n + s}
			\\
			-\Lambda  \left(a \left(-q-r+\sigma _2+\sigma _3\right)+(b+c) \left(p-\sigma _1\right)\right)
			w^{p+q+r-3,\sigma _1+\sigma _2+\sigma _3-4}_{a + b + c, m + n + s} 
			\Bigg)	\\
			-\Lambda  \left(b (r-\sigma _3)-c (q- \sigma _2)\right)
			\Bigg(
			-(m (q+r-2)-(p-1) (n+s)) 
			w^{p+q+r-3,\sigma _1+\sigma _2+\sigma _3-4}_{a + b + c, m + n + s}
			\\
			+\Lambda  \left(a (q+r+1)-a \left(\sigma _2+\sigma _3\right) -(b+c)p+\sigma _1 (b+c)\right)
			w^{p+q+r-2,\sigma _1+\sigma _2+\sigma _3-4}_{a + b + c, m + n + s}
			\Bigg),
		\end{gathered}
	\end{equation}
	and 
	\begin{equation}
		\label{eq:term3}
		\begin{gathered}
			[[w^{r,\sigma_3}_{c,s},w^{p,\sigma_1}_{a,m}],w^{q,\sigma_2}_{b,n}]\\
			=(s(p-1)-m(r-1))\Bigg(
			((q-1) (m+s)-n (p+r-3))
			w^{p+q+r-4,\sigma _1+\sigma _2+\sigma _3-4}_{a + b + c, m + n + s}
			\\
			-\Lambda  \left((a+c) \left(q-\sigma _2\right)+b \left(-p-r+\sigma _1+\sigma _3\right)\right)
			w^{p+q+r-3,\sigma _1+\sigma _2+\sigma _3-4}_{a + b + c, m + n + s} 
			\Bigg)\\
			-\Lambda  \left( c (p-\sigma _1) -a( r- \sigma _3)\right)
			\Bigg(
			((q-1) (m+s)-n (p+r-2))
			w^{p+q+r-3,\sigma _1+\sigma _2+\sigma _3-4}_{a + b + c, m + n + s}
			\\
			-\Lambda  \left((a+c) \left(q-\sigma _2\right)+b \left(-p-r+\sigma _1+\sigma _3-1\right)\right)
			w^{p+q+r-2,\sigma _1+\sigma _2+\sigma _3-4}_{a + b + c, m + n + s}
			\Bigg).
		\end{gathered}
	\end{equation}
	Adding \eqref{eq:term1},\eqref{eq:term2} and \eqref{eq:term3} we get
	 \begin{equation}
	 	\label{eq:jacobi checked}
	 	[[w^{p,\sigma_1}_{a,m},w^{q,\sigma_2}_{b,n}],w^{r,\sigma_3}_{c,s}]+
	 	[[w^{q,\sigma_2}_{b,n},w^{r,\sigma_3}_{c,s}],w^{p,\sigma_1}_{a,m}]+
	 	[[w^{r,\sigma_3}_{c,s},w^{p,\sigma_1}_{a,m}],w^{q,\sigma_2}_{b,n}]=0.
	 \end{equation}
		


\begin{thebibliography}{99}
				
		\bibitem{Strominger:2017zoo}
		A.~Strominger,
		``Lectures on the Infrared Structure of Gravity and Gauge Theory,''
		[arXiv:1703.05448 [hep-th]].
		
		\bibitem{Sorokin:2004ie}
D.~Sorokin,
``Introduction to the classical theory of higher spins,''
AIP Conf. Proc. \textbf{767}, no.1, 172-202 (2005)
doi:10.1063/1.1923335
[arXiv:hep-th/0405069 [hep-th]]

X.~Bekaert, S.~Cnockaert, C.~Iazeolla and M.~A.~Vasiliev,
``Nonlinear higher spin theories in various dimensions,''
[arXiv:hep-th/0503128 [hep-th]].

R.~Rahman and M.~Taronna,
``From Higher Spins to Strings: A Primer,''
Lect. Notes Phys. \textbf{1028}, 1-119 (2024)
doi:10.1007/978-3-031-59656-8{\_}1
[arXiv:1512.07932 [hep-th]].

X.~Bekaert, N.~Boulanger, A.~Campoleoni, M.~Chiodaroli, D.~Francia, M.~Grigoriev, E.~Sezgin and E.~Skvortsov,
``Snowmass White Paper: Higher Spin Gravity and Higher Spin Symmetry,''
[arXiv:2205.01567 [hep-th]].

D.~Ponomarev,
``Basic Introduction to Higher-Spin Theories,''
Int. J. Theor. Phys. \textbf{62}, no.7, 146 (2023)
doi:10.1007/s10773-023-05399-5
[arXiv:2206.15385 [hep-th]].

V.~E.~Didenko and E.~D.~Skvortsov,
``Elements of Vasiliev Theory,''
Lect. Notes Phys. \textbf{1028}, 269-456 (2024)
doi:10.1007/978-3-031-59656-8{\_}3
[arXiv:1401.2975 [hep-th]].

S.~Giombi,
``Higher Spin {\textemdash} CFT Duality,''
doi:10.1142/9789813149441{\_}0003
[arXiv:1607.02967 [hep-th]].

\bibitem{Ponomarev:2016lrm}
D.~Ponomarev and E.~D.~Skvortsov,
``Light-Front Higher-Spin Theories in Flat Space,''
J. Phys. A \textbf{50}, no.9, 095401 (2017)
doi:10.1088/1751-8121/aa56e7
[arXiv:1609.04655 [hep-th]].

\bibitem{Ponomarev:2017nrr}
D.~Ponomarev,
``Chiral Higher Spin Theories and Self-Duality,''
JHEP \textbf{12}, 141 (2017)
doi:10.1007/JHEP12(2017)141
[arXiv:1710.00270 [hep-th]].

\bibitem{Krasnov:2021nsq}
K.~Krasnov, E.~Skvortsov and T.~Tran,
``Actions for self-dual Higher Spin Gravities,''
JHEP \textbf{08}, 076 (2021)
doi:10.1007/JHEP08(2021)076
[arXiv:2105.12782 [hep-th]].

\bibitem{Himwich:2021dau}
		E.~Himwich, M.~Pate and K.~Singh,
		``Celestial operator product expansions and w$_{1+\infty}$ symmetry for all spins,''
		JHEP \textbf{01}, 080 (2022)
		doi:10.1007/JHEP01(2022)080
		[arXiv:2108.07763 [hep-th]].


\bibitem{Monteiro:2022xwq}
R.~Monteiro,
``From Moyal deformations to chiral higher-spin theories and to celestial algebras,''
JHEP \textbf{03}, 062 (2023)
doi:10.1007/JHEP03(2023)062
[arXiv:2212.11266 [hep-th]].	




\bibitem{Ponomarev:2022ryp}
D.~Ponomarev,
``Towards higher-spin holography in flat space,''
JHEP \textbf{01}, 084 (2023)
doi:10.1007/JHEP01(2023)084
[arXiv:2210.04035 [hep-th]].

\bibitem{Ponomarev:2022qkx}
D.~Ponomarev,
``Chiral higher-spin holography in flat space: the Flato-Fronsdal theorem and lower-point functions,''
JHEP \textbf{01}, 048 (2023)
doi:10.1007/JHEP01(2023)048
[arXiv:2210.04036 [hep-th]].
		
\bibitem{Campoleoni:2017mbt}
A.~Campoleoni, D.~Francia and C.~Heissenberg,
``On higher-spin supertranslations and superrotations,''
JHEP \textbf{05}, 120 (2017)
doi:10.1007/JHEP05(2017)120
[arXiv:1703.01351 [hep-th]].


\bibitem{Adamo:2022lah}
T.~Adamo and T.~Tran,
``Higher-spin Yang{\textendash}Mills, amplitudes and self-duality,''
Lett. Math. Phys. \textbf{113}, no.3, 50 (2023)
doi:10.1007/s11005-023-01673-z
[arXiv:2210.07130 [hep-th]].

		
		
		\bibitem{Pasterski:2021rjz}
		S.~Pasterski,
		``Lectures on celestial amplitudes,''
		Eur. Phys. J. C \textbf{81}, no.12, 1062 (2021)
		doi:10.1140/epjc/s10052-021-09846-7
		[arXiv:2108.04801 [hep-th]]
		
		\bibitem{Donnay:2023mrd}
		L.~Donnay,
		``Celestial holography: An asymptotic symmetry perspective,''
		[arXiv:2310.12922 [hep-th]].
		
		
		\bibitem{Pasterski:2016qvg} 
		S.~Pasterski, S.~H.~Shao and A.~Strominger,
		``Flat Space Amplitudes and Conformal Symmetry of the Celestial Sphere,''
		Phys.\ Rev.\ D {\bf 96}, no. 6, 065026 (2017)
		doi:10.1103/PhysRevD.96.065026
		[arXiv:1701.00049 [hep-th]].
		
		S.~Pasterski and S.~H.~Shao,
		``Conformal basis for flat space amplitudes,''
		Phys.\ Rev.\ D {\bf 96}, no. 6, 065022 (2017)
		doi:10.1103/PhysRevD.96.065022
		[arXiv:1705.01027 [hep-th]].  
		
		\bibitem{Banerjee:2018gce} 
		S.~Banerjee,
		``Null Infinity and Unitary Representation of The Poincare Group,''
		JHEP {\bf 1901}, 205 (2019)
		doi:10.1007/JHEP01(2019)205
		[arXiv:1801.10171 [hep-th]].
		
		 \bibitem{Strominger:2013lka} 
A.~Strominger,
``Asymptotic Symmetries of Yang-Mills Theory,''
JHEP {\bf 1407}, 151 (2014)
doi:10.1007/JHEP07(2014)151
[arXiv:1308.0589 [hep-th]].

  A.~Strominger,
  ``On BMS Invariance of Gravitational Scattering,''
  JHEP {\bf 1407}, 152 (2014)
  doi:10.1007/JHEP07(2014)152
  [arXiv:1312.2229 [hep-th]].
  
  \bibitem{Barnich:2009se} 
  G.~Barnich and C.~Troessaert,
  ``Symmetries of asymptotically flat 4 dimensional spacetimes at null infinity revisited,''
Phys.\ Rev.\ Lett.\  {\bf 105}, 111103 (2010)
  doi:10.1103/PhysRevLett.105.111103
  [arXiv:0909.2617 [gr-qc]]. 
 
  T.~He, V.~Lysov, P.~Mitra and A.~Strominger,
  ``BMS supertranslations and Weinberg's soft graviton theorem,''
  JHEP {\bf 1505}, 151 (2015)
  doi:10.1007/JHEP05(2015)151
  [arXiv:1401.7026 [hep-th]].
  
 A.~Strominger and A.~Zhiboedov,
 ``Gravitational Memory, BMS Supertranslations and Soft Theorems,''
 JHEP {\bf 1601}, 086 (2016)
doi:10.1007/JHEP01(2016)086
[arXiv:1411.5745 [hep-th]].


M.~Campiglia and A.~Laddha,
``Asymptotic symmetries and subleading soft graviton theorem,''
Phys. Rev. D \textbf{90} (2014) no.12, 124028
doi:10.1103/PhysRevD.90.124028
[arXiv:1408.2228 [hep-th]].


M.~Campiglia and A.~Laddha,
``New symmetries for the Gravitational S-matrix,''
JHEP \textbf{04} (2015), 076
doi:10.1007/JHEP04(2015)076
[arXiv:1502.02318 [hep-th]].

M.~Campiglia and A.~Laddha,
``Sub-subleading soft gravitons: New symmetries of quantum gravity?,''
Phys. Lett. B \textbf{764} (2017), 218-221
doi:10.1016/j.physletb.2016.11.046
[arXiv:1605.09094 [gr-qc]].


  D.~Kapec, P.~Mitra, A.~M.~Raclariu and A.~Strominger,
  ``2D Stress Tensor for 4D Gravity,''
  Phys.\ Rev.\ Lett.\  {\bf 119}, no. 12, 121601 (2017)
  doi:10.1103/PhysRevLett.119.121601
  [arXiv:1609.00282 [hep-th]].
      
  D.~Kapec, V.~Lysov, S.~Pasterski and A.~Strominger,
  ``Semiclassical Virasoro symmetry of the quantum gravity $ \mathcal{S}$-matrix,''
  JHEP {\bf 1408}, 058 (2014)
  doi:10.1007/JHEP08(2014)058
  [arXiv:1406.3312 [hep-th]].
  
  T.~He, D.~Kapec, A.~M.~Raclariu and A.~Strominger,
  ``Loop-Corrected Virasoro Symmetry of 4D Quantum Gravity,''
  JHEP {\bf 1708}, 050 (2017)
  doi:10.1007/JHEP08(2017)050
  [arXiv:1701.00496 [hep-th]].
  
S.~Banerjee and S.~Pasterski,
``Revisiting the shadow stress tensor in celestial CFT,''
JHEP \textbf{04} (2023), 118
doi:10.1007/JHEP04(2023)118
[arXiv:2212.00257 [hep-th]].


   

  
  


  S.~Stieberger and T.~R.~Taylor,
  ``Symmetries of Celestial Amplitudes,''
  Phys.\ Lett.\ B {\bf 793}, 141 (2019)
  doi:10.1016/j.physletb.2019.03.063
  [arXiv:1812.01080 [hep-th]].
  
M.~Pate, A.~M.~Raclariu, A.~Strominger and E.~Y.~Yuan,
``Celestial operator products of gluons and gravitons,''
Rev. Math. Phys. \textbf{33} (2021) no.09, 2140003
doi:10.1142/S0129055X21400031
[arXiv:1910.07424 [hep-th]].


S.~Banerjee, S.~Ghosh and R.~Gonzo,
``BMS symmetry of celestial OPE,''
JHEP \textbf{04}, 130 (2020)
doi:10.1007/JHEP04(2020)130
[arXiv:2002.00975 [hep-th]].

L.~Donnay, S.~Pasterski and A.~Puhm,
``Asymptotic Symmetries and Celestial CFT,''
JHEP \textbf{09} (2020), 176
doi:10.1007/JHEP09(2020)176
[arXiv:2005.08990 [hep-th]].

\bibitem{Pate:2019lpp}
M.~Pate, A.~M.~Raclariu, A.~Strominger and E.~Y.~Yuan,
``Celestial operator products of gluons and gravitons,''
Rev. Math. Phys. \textbf{33}, no.09, 2140003 (2021)
doi:10.1142/S0129055X21400031
[arXiv:1910.07424 [hep-th]].


  



		
						
		\bibitem{Banerjee:2020zlg}
		S.~Banerjee, S.~Ghosh and P.~Paul,
		``MHV graviton scattering amplitudes and current algebra on the celestial sphere,''
		JHEP \textbf{02} (2021), 176
		doi:10.1007/JHEP02(2021)176
		[arXiv:2008.04330 [hep-th]].
		
		
		
		
		\bibitem{Guevara:2021abz}
		A.~Guevara, E.~Himwich, M.~Pate and A.~Strominger,
		``Holographic symmetry algebras for gauge theory and gravity,''
		JHEP \textbf{11} (2021), 152
		doi:10.1007/JHEP11(2021)152
		[arXiv:2103.03961 [hep-th]].
		
		
		
		\bibitem{Strominger:2021mtt}
		A.~Strominger,
		``$w_{1+\infty}$ Algebra and the Celestial Sphere: Infinite Towers of Soft Graviton, Photon, and Gluon Symmetries,''
		Phys. Rev. Lett. \textbf{127}, no.22, 221601 (2021)
		doi:10.1103/PhysRevLett.127.221601
		
		\bibitem{Adamo:2021lrv}
		T.~Adamo, L.~Mason and A.~Sharma,
		``Celestial $w_{1+\infty}$ Symmetries from Twistor Space,''
		SIGMA \textbf{18}, 016 (2022)
		doi:10.3842/SIGMA.2022.016
		[arXiv:2110.06066 [hep-th]].

		
		\bibitem{Himwich:2023njb}
E.~Himwich and M.~Pate,
``w$_{1+\infty}$ in 4D gravitational scattering,''
JHEP \textbf{07}, 180 (2024)
doi:10.1007/JHEP07(2024)180
[arXiv:2312.08597 [hep-th]].
		
				
		\bibitem{Taylor:2023ajd}
		T.~R.~Taylor and B.~Zhu,
		``w1+{\ensuremath{\infty}} Algebra with a Cosmological Constant and the Celestial Sphere,''
		Phys. Rev. Lett. \textbf{132} (2024) no.22, 221602
		doi:10.1103/PhysRevLett.132.221602
		[arXiv:2312.00876 [hep-th]].
		
		\bibitem{Sheta:2025oep}
A.~Sheta, A.~Strominger, A.~Tropper and H.~Wei,
``Soft Algebras in AdS$_4$ from Light Ray Operators in CFT$_3$,''
[arXiv:2601.00096 [hep-th]].

		
		\bibitem{Strominger:2026yrh}
A.~Strominger and H.~Wei,
``EVERY CFT$_3$ HAS AN $ \mathcal{L}_{\wedge}w_{1+\infty}$ SYMMETRY,''
[arXiv:2603.26459 [hep-th]].
		
		\bibitem{Bittleston:2024rqe}
R.~Bittleston, G.~Bogna, S.~Heuveline, A.~Kmec, L.~Mason and D.~Skinner,
``On AdS$_{4}$ deformations of celestial symmetries,''
JHEP \textbf{07}, 010 (2024)
doi:10.1007/JHEP07(2024)010
[arXiv:2403.18011 [hep-th]].

		
				
				
		\bibitem{deGioia:2023cbd}
L.~P.~de Gioia and A.~M.~Raclariu,
``Celestial sector in CFT: Conformally soft symmetries,''
SciPost Phys. \textbf{17}, no.1, 002 (2024)
doi:10.21468/SciPostPhys.17.1.002
[arXiv:2303.10037 [hep-th]].
		
		\bibitem{Himwich:2025ekg}
E.~Himwich and M.~Pate,
``Light-ray Operators and the ${\rm w}_{1+\infty}$ Algebra,''
[arXiv:2512.18973 [hep-th]].



\bibitem{Ball:2021tmb}
A.~Ball, S.~A.~Narayanan, J.~Salzer and A.~Strominger,
``Perturbatively exact w$_{1+\infty}$ asymptotic symmetry of quantum self-dual gravity,''
JHEP \textbf{01}, 114 (2022)
doi:10.1007/JHEP01(2022)114
[arXiv:2111.10392 [hep-th]].


               \bibitem{Gaberdiel:2017ede}
M.~R.~Gaberdiel and R.~Gopakumar,
``The Higher Spin Square,''
doi:10.1142/9789813144101{\_}0002

		
		\bibitem{Donnay:2018neh} 
		L.~Donnay, A.~Puhm and A.~Strominger,
		``Conformally Soft Photons and Gravitons,''
		JHEP {\bf 1901}, 184 (2019)
		doi:10.1007/JHEP01(2019)184
		[arXiv:1810.05219 [hep-th]].
		
		M.~Pate, A.~M.~Raclariu and A.~Strominger,
		``Conformally Soft Theorem in Gauge Theory,''
		Phys. Rev. D \textbf{100} (2019) no.8, 085017
		doi:10.1103/PhysRevD.100.085017
		[arXiv:1904.10831 [hep-th]].
		
		
		
		D.~Nandan, A.~Schreiber, A.~Volovich and M.~Zlotnikov,
		``Celestial Amplitudes: Conformal Partial Waves and Soft Limits,''
		JHEP \textbf{10} (2019), 018
		doi:10.1007/JHEP10(2019)018
		[arXiv:1904.10940 [hep-th]].
		
		T.~Adamo, L.~Mason and A.~Sharma,
		``Celestial amplitudes and conformal soft theorems,''
		Class. Quant. Grav. \textbf{36} (2019) no.20, 205018
		doi:10.1088/1361-6382/ab42ce
		[arXiv:1905.09224 [hep-th]].
		
		T.~Adamo and T.~Tran,
		``Higher-spin Yang{\textendash}Mills, amplitudes and self-duality,''
		Lett. Math. Phys. \textbf{113} (2023) no.3, 50
		doi:10.1007/s11005-023-01673-z
		[arXiv:2210.07130 [hep-th]].
		
		A.~Puhm,
		``Conformally Soft Theorem in Gravity,''
		JHEP \textbf{09} (2020), 130
		doi:10.1007/JHEP09(2020)130
		[arXiv:1905.09799 [hep-th]].
		
	\end{thebibliography}
\end{document}